# Network strategies to understand the aging process and help age-related drug design

**Gábor I. Simkó, Dávid Gyurkó, Dániel V. Veres, Tibor Nánási and Peter Csermely[1]**

*Semmelweis University, Department of Medical Chemistry, POBox 260, H-1444 Budapest, Hungary*

**Recent studies have demonstrated that network approaches are highly appropriate tools to understand the extreme complexity of the aging process. The generality of the network concept helps to define and study the aging of technological, social networks and ecosystems, which may give novel concepts to cure age-related diseases. The current review focuses on the role of protein-protein interaction networks (interactomes) in aging. Hubs and inter-modular elements of both interactomes and signaling networks are key regulators of the aging process. Aging induces an increase in the permeability of several cellular compartments, such as the cell nucleus, introducing gross changes in the representation of network structures. The large overlap between aging genes and genes of age-related major diseases makes drugs which aid healthy aging promising candidates for the prevention and treatment of age-related diseases, such as cancer, atherosclerosis, diabetes and neurodegenerative disorders. We also discuss a number of possible research options to further explore the potential of the network concept in this important field, and show that multi-target drugs (representing 'magic-buckshots' instead of the traditional 'magic bullets') may become an especially useful class of age-related future drugs.**

## Introduction: Network approaches for the study of the aging process

Aging is one of the most multifactorial, complex processes of living organisms. In spite of this complexity, until very recently the majority of studies examined separate elements of the aging process. The multiplicity of approaches has contributed to the large number of aging theories and definitions.

- According to the antagonistic pleiotropy theory of aging, genes, which are preferable during early development, become detrimental in the aged organism.
- The disposable soma theory of ageing highlights the relocation of resources from somatic maintenance towards increased fertility leading to a slow deterioration of the organism.
- The reliability theory gives a rather descriptive picture of aging emphasizing that aging is a phenomenon of increasing risk of failure with the passage of time.
- The network theory of ageing integrates many elements of the previous theories and describes the shifting balance between various types of damage and repair mechanisms in aging of cellular systems. The young state is characterized by well-repaired damage, while the aged organism can not cope with the accumulated damage and gradually surrenders [1-6].

Aging is accompanied by a number of age-related diseases, cancer, atherosclerosis, diabetes and neurodegenerative disorders, such as Alzheimer's disease, Parkinson's disease or others. For the understanding and complex treatment of these interrelated diseases we need novel approaches.

The network approach proved to be a highly efficient tool to describe complex system behavior including the aging process and age-related diseases. For network description we have to identify separable subsets of the system as network elements, and list their interactions as network contacts or links. Most network elements are also full networks themselves. Thus, elements of social networks, human individuals are networks of organs and cells, cells are networks of proteins, and proteins are networks of amino acids. Networks display a lot of rather general properties, such as

- *small-worldness* providing short pathways between most network elements;
- *scale-free topology* (scale-free distribution of elements having various numbers of neighbors meaning that the probability to find an element with double as many neighbors is halved) enabling the *existence of hubs*, which have a much higher number of neighbors than the average;
- a *community structure* separating the networks to various overlapping groups;

---

[1]*Correspondence: Peter Csermely. csermely@eok.sote.hu*

- co-existence of *strong and weak links*, where the link-strength is usually defined as the real, physical strength of the connection, or as the probability of interactions and
- existence of a *network skeleton*, which is the subset of most important pathways in the network.

Networks provide a framework for the conceptualization of the aging process, but can also be used to understand aging in many ways.
- We may follow changes in the structure of the network during the aging process. One of the possible networks may be the protein-protein interaction network of the cell, where network elements are the cellular proteins and their links represent their physical interactions.
- We may also examine correlation networks, where those elements are connected, which show parallel changes in a given time-period. The elements of various correlation networks may be proteins, genes, but also intracellular organelles, or neurons. Aging typically induces the disorganization of correlated changes, but novel elements of coherent behavior can also be observed. All of these can be nicely demonstrated and analyzed in the network representation.
- Subnetworks may also be defined, where the elements are restricted to only those genes (proteins) that participate in the aging process.

Networks provide a general framework for our understanding of the complexity of life, human conceptualization, culture and technology. It is therefore not surprising that the analysis of the phenomenon of aging can be expanded by the application of a network approach. Table 1 shows a few hierarchical examples of the aging process at different levels from quantum systems, through the more familiar aging of proteins, cells, and organisms, up to the aging of ecosystems (such as forests), social groups (such as the economy) and human technological networks (such as our Windows programs) [6-15]. This generalization of the concept of aging will give us several novel approaches (and repair methods) to understand the aging process better and to cure age-related diseases in entirely novel ways.

**Hiba! A hivatkozási forrás nem található.** Conceptualization of the aging process at the different hierarchical network levels

| Elements of the hierarchical network level | Hallmarks of the aging process |
| --- | --- |
| Elementary particles of quantum systems | Physical equations do not change in time (e.g. the Newton laws did not change in the last few centuries). However, the equations describing a system in a non-equilibrium state do change – this is called an aging process, which is a typical behavior of quantum particles embedded in a thermal bath, or of semi-ordered, glassy materials [7,8]. |
| Monomers of biological macromolecules (amino acids, nucleotides, etc.) | Un-repaired replication, transcription and translation errors accumulate. Various forms of protein (and nucleic acid) damage become more and more prevalent (e.g. in an eighty year old human half of all proteins are estimated to be oxidized), cross-links, occasional proteolytic cuts, amino acid truncations develop [6]. |
| Proteins | Due to the energy loss, and protein damage protein-protein interactions may disappear or loose their affinity. Novel, unexpected, quasi-random protein interactions may also occur. Protein complex composition becomes 'wobbly', fuzzy. Proteins are dislocated and appear in unusual cellular compartments [9,10]. |
| Cells | Inter-cellular interactions may become irreversibly tight (e.g. by developing cross-links) or too loose, gradually loosing their high-affinity contacts. Since the development of inter-cellular contacts is costly, functional brain networks show loss of their small-world properties in the age-related Alzheimer's disease [11]. |
| Organisms | The social network of aged individuals usually deteriorates, shrinks, keeping only the most important contacts for major remaining social functions. This contributes to the age-related cognitive decline and to the loss of emotional support leading to increased frailty [12]. Ecosystems, like forests also show the hallmarks of aging [13]. |
| Social groups | A network of social groups, such as a network of firms may also display the signs of aging as it has been shown in the declining network of the New York garment industry by Brian Uzzi and colleagues [14]. |
| Ecosystems forming a global ecological network | The aging research of the global ecosystem of Earth, Gaia, is in its infancy at the moment. However, our increasingly integrated knowledge, such as the global river network [15] give us more and more tools to assess the rather worrisome aging of our habitat. |
| Elements of human conceptual, cultural and technological systems | Human conceptual networks (such as arguments over a complex issue; cross-references in textbooks, etc.), cultural networks (such as the network of actors in a Shakespeare drama, movie actor networks, etc.), or technological networks (such as electric power grids, computer programs, the Internet, etc.) may also show typical signs of aging. As a trivial example we all experience more frequent errors of our Windows program- |

| | network when the system gets older. |
|---|---|

Since the creation and maintenance of newly established links between two network elements are costly, and aging is usually accompanied by a decline in the system's resources, aging – generally – induces a loss of links, leading to a declining network. However, aging is also hallmarked by the emergence of non-specific contacts due to the impaired recognition or positioning of potential partners. The increase of non-specific contacts leads to the appearance of novel links within the network, which affect the global network topology. Small-worldness is often lost during aging (distant elements can not find each other so easily in an aged network), and many times elements with a large number of contacts (hubs) vanish or just inversely: specific, age-related hubs appear [6,10].

Link-removal and link-appearance are continuous, parallel events in the dynamic life of most networks. Somewhat surprisingly their balance does not usually result in a 'mixed' network structure, but rather results in two distinguishable network forms. One of them is the 'forever-young', 'ageless', r-strategist-like (proliferation-optimized) network, where link-formation is prevalent, the overlap of network communities (modules) is high, and the structure is flexible. The other prevalent type is the 'always old', overspecialized, K-strategist-like (survival-optimized) network, where link-decay is most common, the overlap of network communities is small, and the structure is rigid [16]. The shift of topology towards a less overlapping, more rigid structure might by itself suggest the aging of the system described by a given network.

Networks do not only age themselves, but also affect the aging of their components. The network context may slow down the aging of network elements, exemplified by the effect of social networks to elderly people [12]. On the contrary, networks may also channel unexpected damage, leading to an accelerated aging of their components, in a similar way as the avalanche of mitochondria-related free radicals accelerates the aging of all molecules around [17].

Cellular networks may be divided into six categories [6]: structural networks which include protein-protein interactions, cytoskeletal and membrane-organelle networks, and three functional networks: gene transcription, signaling and metabolic networks. Our review focuses on the protein-protein interaction networks among these cellular networks.

## Protein-protein interaction networks in aging

In protein-protein interaction networks (interactomes) elements are proteins and the links between them are their physical interactions. Ideally, the weight of a protein-protein interaction link should correspond to the affinity (strength) of the physical interaction between the two proteins. However, current protein-protein interaction databases use agglomerated data, which conceptualize the weight as the probability of the actual interaction and average the important effects of simultaneous expression, intracellular compartmentalization, posttranslational modifications, etc. Due to this probability function-like network concept, the links of protein-protein interaction networks usually do not have directions. Signaling networks may be considered as subnetworks of protein-protein interaction networks, where the elements (signaling molecules) represent a segment of the proteins participating in the complete interactome, and are connected with directed and (so-called) colored links, where colors show if the interaction has an activating or inhibiting role.

The protein-protein interaction network of aging-associated genes (or longevity networks as referred to Budovsky *et al*. [18]) is another subnetwork of the interactome containing the proteins of aging-related genes [19] as well as the links between them. Figure 1 shows the human protein-protein interaction network of aging-associated genes [19-21]. Importantly, the network is a continuous network (only the GHRH and GHRHR aging-associated proteins are not participating in its giant component shown on the figure). The extensive coverage of age-related genes by the giant component of the related longevity network has also been shown by Budovsky *et al*. [18]. The network shown on Figure 1 includes a large number of key signaling proteins. P53 and GRB2 are two prominent hubs having a large number of neighbors and occupying a central position in the network characterized by a high 'betweenness centrality'. P53 is a transcription factor that plays a prominent role in the regulation of the cell cycle, stress response, apoptosis and has both aging and anti-aging effects [22]. GRB2 is an important adaptor protein in growth signaling, although its direct role in aging has not been elucidated yet. Members of the MAPK/ERK, PI3K/AKT and JAK/STAT pathways are between the 20 largest hubs of the network (Figure 1). These pathways occupy key positions of the network suggesting that they have a high impact on aging. The importance of hubs in age-related gene networks was also demonstrated by earlier studies of Promislow [23], Ferrarini *et al*. [24], Budovsky *et al*. [18] and Bell *et al*. [25]. The overrepresentation of signaling proteins was also noted by Wolfson *et al*. [26], who showed that a common signaling signature network of human longevity and major age-related disease genes exists, and that it includes the insulin pathway, and – somewhat surprisingly – the adherens junctions- and focal adhesion-related signaling.

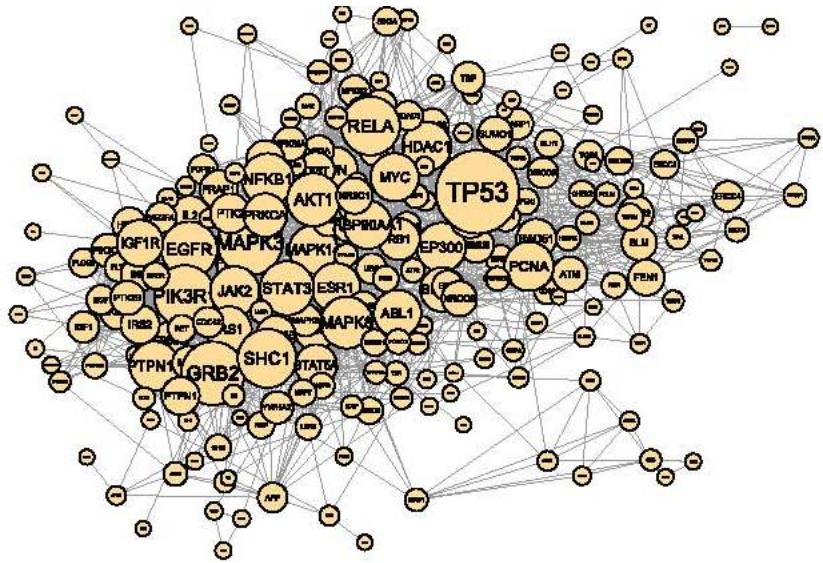

**Figure 1** The human protein-protein interaction network of aging-associated genes. 261 aging-associated genes were assembled using the GenAge Human Database [19]. Protein-protein interactions of the human interactome were collected from the 8.0 version of the STRING Database [20] using physical contacts only. The network was visualized using the Cytoscape program [21]. The degree (number of neighbors) of nodes is represented by the size of the circle and the font. Note the high number of signaling pathway proteins among hubs (nodes with degrees – and therefore size – much greater than average), exemplified by the MAPK/ERK and PI3K/AKT proteins.

Besides the subnetwork of the aging-related gene protein products, a more extended network including their neighbors and the neighbors' connections can also be created [18,25,26]. This extended network, contains a lot of proteins, the neighbors of age-related proteins, which have not been recognized yet as aging-related genes. However, these proteins give an excellent target-set to identify further aging-related genes, as it has been demonstrated by Bell *et al.* [25].

Xue *et al.* [27] went yet a step further, and examined changes in the whole fruit fly and human brain interactome during aging. They showed that age-related changes preferentially affect only a few modules (communities, groups) of the network, and aging-associated proteins are typically located between modules, providing an important element of the regulation of network functions. Molecular chaperones also have a preferential inter-modular localization. Their inter-modular position is special, since they link distant network modules with low-affinity, weak links. This 'chaperone-behavior' stabilizes the network [10]. On the other hand, chaperones constitute an important repair mechanism that slows down the aging process. During stress chaperones become occupied by damaged proteins, which contributes to a larger separation of network modules [6,28]. We expect that the same mechanism occurs during the aging process. Chaperones and presumably a large number of other proteins (in yeast this class constitutes 5% of the whole genome [29]) lose their module-connecting, stabilizing role in aging cells and organisms. The disappearing weak-links destabilize the network, the modules fall apart, their regulation becomes deteriorated, and the separated modules cannot optimally fulfill their tasks. Such changes lead to increased noise and destabilization in the network, which correspond well with the typical signs of aging.

The GenAge Database [19] contains almost three hundred aging-associated human genes. Most of these genes are not acting alone, but as a part of many smaller and larger protein complexes. Currently, the complexity of the age-related protein-protein interactions is revealed most by the cross-talks of age-related signaling pathways. As an example, the growth hormone-related pathways, the oxidative stress-induced pathway and the dietary restriction pathway all affect the FOXO (Daf-16) transcription factor [30]. Many more focal points of age-related signaling may emerge in the future. Network analysis will certainly offer a great help in the identification of these key elements, which may also serve as drug target candidates in promotion of healthy aging and combating age-related diseases.

Despite of our enthusiasm, we must note that the network approach is just at its infancy in aging research. Thus, we can not list – yet – those genes, which have been predicted by network-related methods as participants in aging and verified later experimentally. However, we may already highlight a number of candidates for novel longevity-genes as neighbors of age-related proteins. Among these candidates those are promising especially, which are neighboring hubs with high centrality in the network. Moreover, with modern network prediction methods [31] we may also predict those links and network elements, which should be a part of the network, but have not been identified yet. These network extensions may also give us a number of research-targets for future age-related studies. Network-based analysis may also give a molecular-level explanation of age-related macroscopic features such as the increased unpredictability, diversity and destabilization of aged organisms including elderly people. Such features are so-called 'emergent properties' of networks, which are displayed by the whole network, but can not be predicted from any single network element. Aging is characterized by a large number such 'emergent-like' properties, where we need the understanding of the whole cellular system and not only its details. Networks provide a very efficient tool for accomplishing this goal.

**Age-related changes in cell compartmentalization**
As we have shown before, the signaling subnetwork becomes an especially important part of protein-protein interaction networks in the elucidation of the mechanisms of aging. In this section we will consider the involvement of various cell compartments (such as the nucleus, endoplasmic reticulum, Golgi, mitochondria, plasma membrane and cytoplasm) in signaling processes. Figure 2 shows the age-related signaling network of cellular compartments, where network elements are the cellular compartments and links between two compartments represent various ageing-related signaling pathways [30] involving both compartments. The major cell compartments are contributing to the age-related signaling pathways to a similar extent with the exception of the Golgi apparatus. This large spatial complexity of signaling events is in agreement with the multifactorial nature of age-related signaling. In the network representation mitochondria and the endoplasmic reticulum are tightly coupled, which is in agreement with the results of cell biological studies.

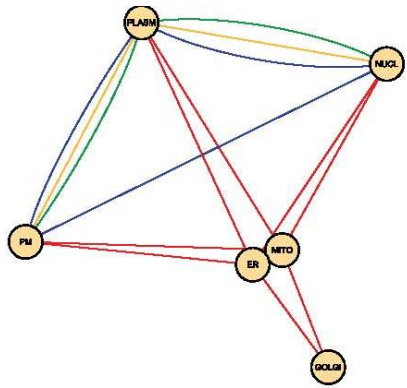

**Figure 2** Age-related signaling network of cellular compartments. In this network representation, the elements of the network are the cellular compartments (PM, plasma membrane; ER, endoplasmic reticulum; MITO, mitochondria; GOLGI, the Golgi apparatus; NUCL, nucleus and PLASM, cytoplasm), while the links between them represent the age-related signaling pathways [30]. The network has been visualized using the Cytoscape program [21]. The colors represent the following pathways: orange – growth hormone pathways; green – MAP-kinase cascade; blue – dietary restriction pathway; red – reactive oxygen species, ROS.

During the aging process the nuclear pore complexes become more permeable [9]. This key element of cellular aging not only disturbs the nuclear reassembly after mitosis and compromises nuclear integrity exposing DNA to oxidative damage, but may also significantly disturb all nucleus-related signaling steps, including the growth hormone, insulin, TGF-ß, dietary restriction, and oxidative stress mediated signaling pathways.

Other cellular compartments, such as mitochondria, the endoplasmic reticulum or the plasma membrane may also become more permeable in aged cells. Indeed, an increased susceptibility of the mitochondrial permeability transition pore has been reported in aged mice [32] and in interfibrillar heart mitochondria of aged rats. The age-related reduction of $Ca^{2+}$ retention capacity of interfibrillar heart mitochondria may explain the increased susceptibility to stress-induced cell death in the aged myocardium [33]. Similarly, an age-dependent decline of sarcoplasmic reticulum ultrastructure has been reported leading to irregular $Ca^{2+}$ signaling. This structural change is related to the decreased expression of the mitsugumin-29 protein [34]. The increased fragility of the plasma membrane of senescent cells may be a result of increased exosome formation in the senescent state, as seen in normal and prostate cancer cells [35]. An increased exosome formation also accompanies the development of drusen, an extracellular deposit representing a significant risk factor of age-dependent macular degeneration [36].

## Age-related drug development

Currently, there is only a small number of drugs on the market, which directly target the aging process. Most of the available medicaments prevent skin-aging, such as a kinetin-based drug, which delays the effects of aging in human skin fibroblasts [37], or an algae-extract, which activates the proteasome, and delays the aging of human keratinocytes [38]. The skin rejuvenating action of several other compounds (e.g. ethanolamine [39], or 4-oxo-retinol [40]) is patented, although their mechanism of action and biological importance has not yet been fully established. A recent study showed that rapamycin treatment started at already aged mice (at 600 days of life) extended both the median and maximal lifespan of the animals [41].

Several studies [42-44] showed that the development of multi-target drugs might give better results than the traditional 'magic bullets' targeting a single protein. Single target-design might not always give satisfactory results, as there might be a backup system, which replaces the functions of the inhibited target protein. The low-affinity binding of multi-target drugs increases the druggable proteome (i.e. the number those proteins, which are able to bind a drug-like, small and hydrophobic, orally administered compound with a reasonably high affinity), hence the number of potential drug targets. Multiple targeting allows lower doses that often result in less side effects, toxicity and resistance. By using multi-target drugs we can decrease the functionality of entire protein cascades, which explains how multi-target design may produce more effective results, while not influencing drastically any components within the system.

The network-approach opens several ways to design a multi-target drug. One may attack hubs of the protein-protein interaction network, 'hub-links' (links connecting hubs), bridges (inter-modular links having a central position characterized by a high 'betweenness centrality') or elements in the overlaps of numerous network modules [43]. Moreover, perturbation analysis of protein-protein interaction networks may highlight those alternative target-sets for multi-target drug design, where the initial effects (e.g. mutations or protein damage inducing an age-related disease or aging) accumulate their actions [45].

Recent studies have shown that aging is strongly linked with age-related diseases, and they share a common signaling network [18,26]. Signaling hubs of the age-related protein-protein interaction subnetwork may be good candidates for age-related drug-targets, which may also help to prevent or cure age-related diseases. Moreover, the extended age-related protein-protein interaction network including the neighbors of ageing gene products contains a large number of yet unknown age-related proteins that offer additional drug-target options [25,26]. The central position of key age-related proteins in the interactome raises the possibility that appropriately selected subsets may form efficient target-sets for multi-target drug design opening a way to ensure healthy ageing instead of combating each age-related disease one by one.

## Conclusions and perspectives

In the introduction of the current review we showed that the network approach can grossly expand the aging concept from quantum systems up to the complex ecosystem of the Earth (Table 1). We are at the very beginning of both the conceptualization and studies of the aging process of the internet, world-wide-web, social networks, forests, trade and co-ownership networks of the economy, etc. This network-related generalization of the aging concept will make a large unexpected benefit allowing us to understand the aging and age-related diseases of our body from a number of entirely novel perspectives. The use of the dual concept of 'forever young' (flexible, proliferation/exploration-optimized) and 'always-old' (rigid/overspecialized, survival-optimized) networks [16] will certainly help the generalization of our biological and medical knowledge on aging of other complex systems.

Our review focuses on the role of protein-protein interaction networks (interactomes) and their subnetworks, signaling networks in aging. The highlights of the current studies can be summarized as follows. Hubs and inter-modular elements of both protein-protein interaction and signaling networks proved to be of great importance in the regulation of the aging process. It is a great challenge of further studies to enrich this list with other topologically or dynamically important network positions, such as that of creative network elements [46]. The thorough analysis of age-related network positions will increase the accuracy of the prediction of novel age-related genes. Signaling proteins are highly over-represented in age-related gene products including network hubs. The age-related signaling network components can also be found among the major age-related disease genes forming a common signaling signature.

Aging induces a rather general permeability-increase of various cellular compartments, such as the nucleus, the endoplasmic reticulum and mitochondria. However, this increase is most probably not a consequence of increased membrane flexibility. Just the contrary, the increased permeability (or susceptibility for permeability-increase) may reflect an increased membrane rigidity causing membrane fragility. Changes in the permeability of cellular compartments certainly rearrange the actual representation of both interactomes and signaling networks. These assumptions open a number of exciting areas for further studies.

All the above findings are good starting points to find novel drug targets helping healthy aging and extending the healthy lifespan. Multi-target drugs will be especially helpful to address the extreme complexity of aging. The large overlap between network components participating in the regulation of the aging process and in age-related major diseases, such as cancer, atherosclerosis, diabetes, or neurodegenerative diseases, makes the development of age-related multi-target drugs especially promising, since by the help of these 'magic buckshots' (instead of the traditional 'magic bullets') we may help the prevention and targeting of many age-related diseases altogether.

**Abbreviations:** GHRH and GHRHR, growth-hormone-releasing hormone and its receptor; MAPK/ERK, the mitogen-activated protein kinase/extracellular signal-regulated kinase, 'classical' MAP kinase signaling pathway; JAK/STAT, the Janus kinase/signal transducers and activators of transcription protein signaling pathway; PI3K/AKT, the phosphoinositide-3-kinase/protein kinase B signaling pathway; TGF-ß, transforming growth factor beta.

**Competing interests :** The authors' declare that they have no competing interests.

**Authors' contribution:** GS made the manuscript figures, wrote a significant part of the protein-protein interaction network and drug-development sections and helped to finalize the manuscript; DG had important contributions to the introduction, Table 1 and the finalization of the manuscript; DV and TN contributed to Figure 2 and to the cell compartmentalization section; PC made the outline and the integration of the manuscript.

**Authors' information:** All authors are members of the LINK-Group (www.linkgroup.hu). GS and DG are completing their MSc theses, DV and TN are undergraduate research students and PC is a professor of biochemistry and network studies.

**Acknowledgements:** Work in the authors' laboratory was supported by the EU (FP6-518230) and the Hungarian National Science Foundation (OTKA K69105).